# Couple Control Model Implementation on Antagonistic Mono- and Bi-Articular Actuators *

Flavio Prattico, Mohd Azuwan Mat Dzahir, and Shin-ichiroh Yamamoto

*Abstract*— Recently, robot assisted therapy devices are increasingly used for spinal cord injury (SCI) rehabilitation in assisting handicapped patients to regain their impaired movements. Assistive robotic systems may not be able to cure or fully compensate impairments, but it should be able to assist certain impaired functions and ease movements. In this study, a couple control model for lower-limb orthosis of a body weight support gait training system is proposed. The developed leg orthosis implements the use of pneumatic artificial muscle as an actuation system. The pneumatic muscle was arranged antagonistically to form two pair of mono-articular muscles (i.e., hip and knee joints), and a pair of bi-articular actuators (i.e., rectus femoris and hamstring). The results of the proposed couple control model showed that, it was able to simultaneously control the antagonistic mono- and bi-articular actuators and sufficiently performed walking motion of the leg orthosis.

## I. INTRODUCTION

In this research, a static characterization of a new pneumatic muscle was conducted in order to model analytically its behaviour. The main task is to find, as made by [1], the force that the pneumatic muscle can apply as a function of the supply pressure and of its contraction. It is possible to see the experimental characterization, particularly by maintain fixed the pneumatic muscle at various contraction steps and, varying the supply pressure, then the force that it produces was recorded. We choose to not use, in our system, the pneumatic muscle in the negative part of the contraction, and then we not characterize it there. In addition, it is possible to note, that the main properties of the pneumatic muscle are very similar to those of the commercial pneumatic muscles. The successive step of this analysis is to fit with a surface of the previous characterization. We choose to fit the surface with a two variables polynomial function. We need to express the supply pressure as a function of the force and the contraction. To do this the fitting equation must be solvable in the term of the pressure, then the term of the pressure must have a degree equal or less to two (different approach used in [1] in which the equation is fifth degree in both variables, then needs to solve numerically with long time of computing). We then conducted a sensibility analysis on the degree of the fitting equation. Particularly we compute the Root Mean Square Error (RMSE) between the experimental point and the fitting surface and we express the results as a function of the degrees of the two variables x and y. The results of the characterization are summarized in the Table 1. It is possible to note that, we have a great reduction of the RMSE from first to second degree in x and, at the same time, we choose to have third degree in y. This choice is due to the fact that we do not have a great reduction of the RMSE between third and fourth degree in y and then we decide to reduce the number of the parameters to increase the computation speed. The resulting fitting equation is:

$$f(x,y) = a_1 + a_2 x + a_3 y + a_4 x^2 + a_5 xy + a_6 y^2 + a_7 x^2 y$$
$$+ a_8 xy^2 + a_9 y^3 \quad ...(1)$$

Where, x represents the supply pressure, y is the contraction and f(x; y) is the force.

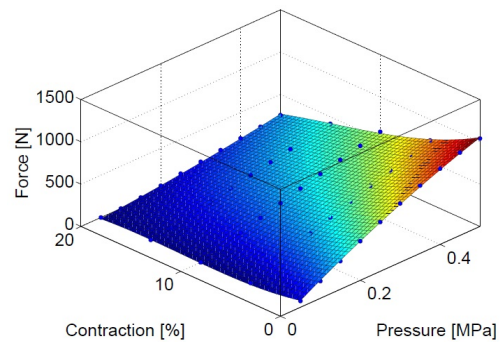

Figure 1: Graphical visualization of the fitting polynomial equation (blue dot are the experimental points)

Table 1: Sensibility analysis

| Degree of $x$ | Degree of $y$ | RMSE |
|---|---|---|
| 1 | 1 | 88.2 |
| 2 | 1 | 36.7 |
| 2 | 2 | 23.2 |
| 2 | 3 | 11.6 |
| 2 | 4 | 7.4 |
| 1 | 2 | 24.3 |
| 1 | 3 | 18.4 |
| 1 | 4 | 17.2 |

*Research supported by KAKENHI.

Flavio Prattico, He received master's degree (2011) in Mechanical Engineering from University of L'Aquila. Since Oct. 2011, he has been a doctoral course at University of L'Aquila, Department of Mechanical, Energetic, and Processing Engineering.

Mohd Azuwan Mat Dzahir, He received master's degree (2011) in mechanical engineering from Faculty of Mechanical Engineering at Universiti Teknologi Malaysia; UTM. Since Sept. 2011, he has been a doctoral course at SIT. 307 Fukasaku, Minuma-ku, Saitama-City, Saitama, 337-8570 Japan. (phone:+080-4094-8009;e-mail: azuwan@fkm.utm.my).

Shin-ichiroh Yamamoto, He received the Ph.D. (2000) degrees in science from the Department of Life Science, The University of Tokyo. He is a professor at SIT, Japan. (phone:+81-48-720-6024; e-mail: yamashin@se.shibaura-it.ac.jp).

## II. OVERVIEW OF AIRGAIT EXOSKELETON'S LEG ORTHOSIS

Figure 2 shows the AIRGAIT exoskeleton's leg orthosis of the developed body weight support gait training system used for this research. The leg orthosis system implemented six pneumatic muscles which antagonistically arranged based on the human musculoskeletal system (i.e., mono- and bi-articular muscles).

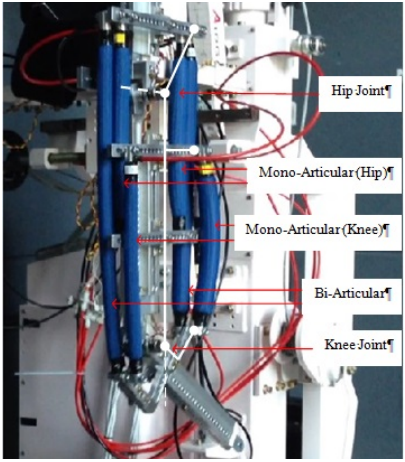

Figure 2: Overview of AIRGAIT exoskeleton's leg orthosis.

The pneumatic muscle used in this research is a self fabricated McKibben artificial muscle actuator. The input pressure of the pneumatic muscle is regulated by electro-pneumatic regulator separately for each actuator. The special characteristic of pneumatic muscle will cause it to contract when the air pressure is supplied, and expand when the air pressure is removed. In other words, the pneumatic muscle is able to emulate the force and muscle contraction of human's muscle. In addition, it is also might be able to perform similar contractions and expansions, where their movement is almost similar to the movements of the human's muscles. The measurement of the joint excursions (i.e., hip and knee) is made using potentiometer. This system uses the Lab-View software and RIO module to provide the input signals and to read the output data of the leg orthosis.

## III. METHODS

To determine the required supply pressure as a function of a force and contraction, a fitting equation was first characterized using the polynomial function. Then, the sensibility test was carried out using the RMSE between the experiment points and fitting surface to determine the highest orders of variable x and y. This will resulted to a good reduction in the RMSE value. Couple control was designed using the Newton Euler equation model. Based on this model, the allocated force requires for agonist and antagonist actuators could be determined by introducing a joint's stiffness. By using a simple approach, the mono- and bi-articular actuators model as a function of joint's angle was designed to estimate the agonist and antagonist co-contraction's value. Two tests were performed; the first is using the co-contraction model of position control [3]; and the second is using the designed couple control. Both tests were evaluated and compared at frequency of 0.5Hz (2s gait cycle speed).

## IV. CONTRACTION MODEL OF ANTAGONISTIC MONO- AND BI-ARTICULAR ACTUATORS

Figure 3 and 4 show the mono- and bi-articular actuators model's contraction. Based on the information gained from the AIRGAIT exoskeleton's leg orthosis, the contraction of the antagonistic mono- and bi-articular actuators are derived using the trigonometric function. The antagonistic mono-articular actuator's contraction for the hip joint ($\theta_1$) is:

$$AB = \sqrt{\overline{AC}^2 + \overline{BC}^2 - 2\overline{AC}.\overline{BC}cos(\alpha)} \quad ...(2)$$

$$\theta_0 = \alpha|_{\theta_1=0}$$

$$Antagonist: \alpha = \theta_1 - \theta_0 = \alpha(\theta_1)$$

$$Agonist: \alpha = \theta_1 + \theta_0 = \alpha(\theta_1)$$

$$k(\theta_1) = k_1^{ant.} = k_1^{ag.} = \frac{l_m - AB}{l_m} \quad ...(3)$$

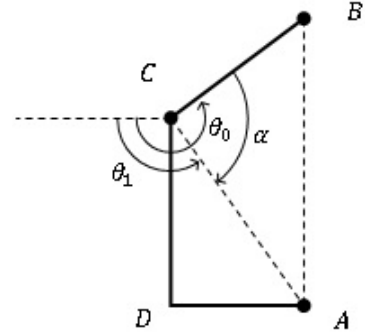

Figure 3: Mono-articular actuator model's contraction.

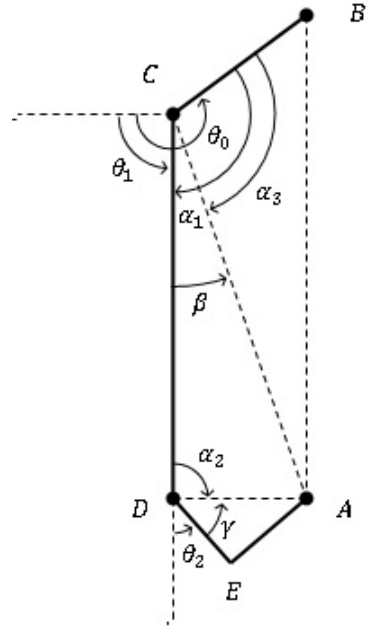

Figure 4: Bi-articular actuator model's contraction.

The implementation is also similar for the antagonistic mono-articular actuators for the knee joint ($\theta_2$). However, the axis for the knee joint motion is based on the line formed between the hip and knee joints. The antagonistic bi-articular actuator's contraction is derived as follows:

$$AC = \sqrt{\overline{AD}^2 + \overline{CD}^2 - 2\overline{AD}.\overline{CD}\cos(\alpha_2)} \quad ...(4)$$

$$\alpha_2 = 180 - \gamma - \theta_2$$

$$AB = \sqrt{\overline{AC}^2 + \overline{CB}^2 - 2\overline{AC}.\overline{CB}\cos(\alpha_3)} \quad ...(5)$$

$$\alpha_3 = \alpha_1 - \beta, \quad \alpha_1(\theta_1) = \theta_0 - \theta_1$$

$$\beta = a\cos\left(\frac{\overline{CD}^2 + \overline{AC}^2 - \overline{DA}^2}{2.\overline{CD}.\overline{AC}}\right) \quad ...(6)$$

$$k(\theta_1, \theta_2) = \frac{l_m - AB}{l_m} \quad ...(7)$$

## V. Newton-Euler Equation Model

The Newton-Euler equation model could be determined once we have calculated the geometric parameters such as masses, inertias and dimensions. The couples $C_1$ and $C_2$ at the joints are determine as follows:

$$C_1 = I_{11}\ddot{\theta}_1 + I_{22}\ddot{\theta}_1 + I_{22}\ddot{\theta}_2 + \ddot{\theta}_1 d_1^2 m_1 + \ddot{\theta}_1 d_2^2 m_2 + \ddot{\theta}_2 d_2^2 m_2$$
$$+\ddot{\theta}_1 d_{21}^2 m_2 + d_{21} g m_2 \cos(\theta_1 + \theta_2) + d_1 g m_1 \cos(\theta_1)$$
$$+d_{21} g m_2 \cos(\theta_1) - \dot{\theta}_2^2 d_2 d_{21} m_2 \sin(\theta_2)$$
$$+2\dot{\theta}_1^2 d_2 d_{21} m_2 \cos(\theta_2) + \dot{\theta}_2^2 d_2 d_{21} m_2 \cos(\theta_2)$$
$$-2\dot{\theta}_1\dot{\theta}_2 d_2 d_{21} m_2 \sin(\theta_2) \quad ...(8)$$

$$C_2 = I_{22}(\ddot{\theta}_1 + \ddot{\theta}_2) + d_2 m_2 \begin{pmatrix} d_{21}\sin(\theta_2)\dot{\theta}_1^2 \\ +g\cos(\theta_1 + \theta_2) \\ +d_2(\ddot{\theta}_1 + \ddot{\theta}_2) \\ +\ddot{\theta}_1 d_{21}\cos(\theta_2) \end{pmatrix} \quad ...(9)$$

## VI. Control Model

There are in literature a lot of control model, they can be divided into two main groups: position and couple control. The last one requires a completely description of the system and, if it has a high number of degree of freedom, the formulation of the couple joint expression become difficult. For this reason many of the controllers used in industry are based on empirical approach as the fuzzy or the PID controls. The main idea of the proposed model is based on the control of the position by controlling the couples of the joints and varying the stiffness of the system as a function of the degree of precision required and of moving masses. The model is composed of a part that describe the geometric configuration between the pneumatic muscles and the joints, another part for the computing of the joints couple based on the Newton-Euler equations and the last part that able us to compute the needed supply pressure knowing the equivalent forces and the contractions. First of all, we can define the stiffness of a system as the measure of the resistance to the deformations. For our system this concept of stiffness translates itself into the level of the force of the antagonist muscle that we can call the "base force" (following a similar nomenclature proposed by [2]). In order to describe the control model we can set and define, for the moment, as R = cost the stiffness of the system that represent the force of the antagonist pneumatic muscle. From the geometrical model we can find the percentage contraction of the two muscles as a function of the angle θ, then:

$$k_1^{ag.} = f(\theta_1) \text{ and } k_1^{ant.} = f(\theta_1) \quad ...(10)$$
$$k_2^{ag.} = f(\theta_2) \text{ and } k_1^{ant.} = f(\theta_2) \quad ...(11)$$
$$k_3^{ag.} = f(\theta_1, \theta_2) \text{ and } k_3^{ant.} = f(\theta_1, \theta_2) \quad ...(12)$$

Where $k_1^{ag}$ represents the contraction of the agonist muscle of the joint 1, instead $k_2^{ant}$ is the contraction of the antagonist muscle of the joint 2. From the NE equations we can compute the couples C1 and C2 as follow:

$$C_1 = f(m_1, I_1, \theta_1, \dot{\theta}_1, \ddot{\theta}_1) \quad ...(13)$$
$$C_2 = f(m_2, I_2, \theta_2, \dot{\theta}_2, \ddot{\theta}_2) \quad ...(14)$$

But geometrically the couples C1 and C2 can be also computed as:

$$C_1 = (F_1^{ag.} - F_1^{ant.}).l_1, \quad C_2 = (F_2^{ag.} - F_2^{ant.}).l_2 \quad ...(15)$$

Being $F_1^{ant}$, and $F_2^{ant}$, equal to R, from the last equations we can compute the forces $F_1^{ag}$, and $F_2^{ag}$:

$$F_1^{ag.} = \frac{C_1}{l} + R, \quad F_2^{ag.} = \frac{C_2}{l} + R \quad ...(16)$$

Now we have to find the pressures that correspond to the forces $F_1^{ag}$ and $F_2^{ag}$. To do this it is necessary to invert the equation of the fit of the pneumatic muscle characterization. The equation showed in the previous section is of the second degree in x and then we can solve it easily:

$$A = a_4 + a_7 y \quad ...(17)$$
$$B = a_2 + a_5 y + a_8 y^2 \quad ...(18)$$
$$C = a_1 + a_3 y + a_6 y^2 + a_9 y^3 - f(x, y) \quad ...(19)$$
$$x = \frac{-B}{2A} \pm \frac{\sqrt{B^2 - 4AC}}{2A} \quad ...(20)$$

And considering the physical meaning of x, y and f(x; y), the equation can be summarize as P = f(F;K). Then, known the force and the percentage contraction of the muscle we can compute the pressure:

$$P_1^{ag.} = f(F_1^{ag.}, k_1^{ag.}) \text{ and } P_2^{ant.} = f(F_2^{ant.}, k_2^{ant.}) \quad ...(21)$$

## VII. Results and Discussions

Figure 6 shows the results of co-contraction based position control and couple control of the AIRGAIT exoskeleton leg orthosis using both antagonistic mono- and bi-articular actuators. The result showed that, at a gait cycle speed of 0.70m/s, both control model schemes were able to give a sufficient hip joint excursion of the leg orthosis walking motion. An approximately only 0.2s time shift could also be seen in the joint excursion graphs. However, the developed couple control was proved to be much effective to control the antagonistic actuators of the leg orthosis at the evaluated gait cycle speed when compared to the co-contraction based position control alone. This might be because of the simultaneous movements of all antagonistic mono- and bi-articular actuators (total of six) which are limiting the operating gait cycle speed when using the position control.

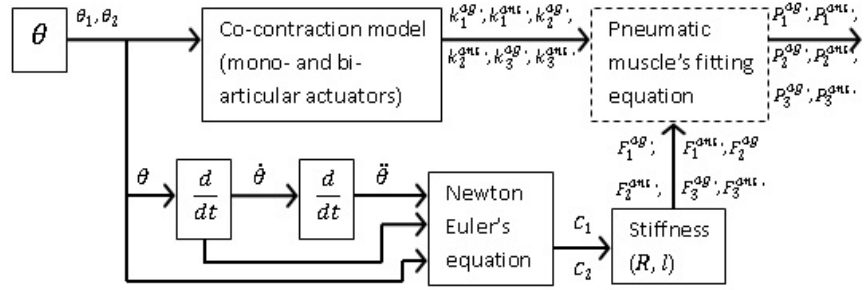

Figure 5: Couple control model.

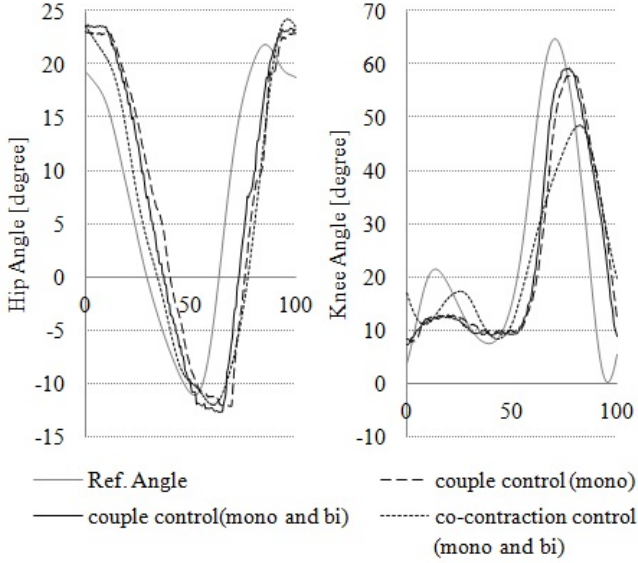

Figure 6: Hip and knee joint excursions of couple control model for AIRGAIT exoskeleton's leg orthosis.

The similar pattern could also been shown in the previous design of AIRGAIT exoskeleton's leg orthosis. Where, the performance of the leg orthosis control dropped at 2s (0.70m/s) gait cycle when the operating gait cycle speed was increased from 5s (0.28m/s) up to 1s (1.40m/s) gait cycle [3]. The implementation of couple control which considers the inertia model of leg orthosis as well as contraction movements of antagonistic mono- and bi-articular actuators was able to comprehend the simultaneous movement of all antagonistic actuators at evaluated gait cycle speed of 0.70m/s. This showed that, a consideration of inertia model was also played an important role in precisely control the movements of all six antagonistic mono- and bi-articular actuators simultaneously along with the co-contraction movement of the antagonistic actuators. The implementation of antagonistic mono-articular actuators alone using the developed couple control was also been performed to evaluated the importance of the addition of antagonistic bi-articular actuators into the design of the leg orthosis. A slightly improvement in time shift and maximum joint excursion could be seen in the graphs. However, a test with subject has yet to be performed. The result with a subject might give a much evident difference between the control of leg orthosis with antagonistic mono-articular actuators alone and with addition of bi-articular actuators.

## VIII. CONCLUSION

In this paper, couple control model was designed to improve the previous designed co-contraction based position control model. These control models were designed to actuate the antagonistic mono- and bi-articular actuators of the AIRGAIT exoskeleton's leg orthosis. In the previous papers, the co-contraction based position control model was proved to be efficient in actuating the antagonistic actuators. However, there is a limitation in the maximum operating gait cycle speed. Therefore, the couple control was introduced to improve the performance of the leg orthosis control. This is because; the performance drop of the leg orthosis control was caused due to the inertia and gravitational effects of the leg orthosis. The results showed that, the introduction of the inertia model into the control system was able to improve the operating speed of the control system. However, this result of the couple control could be improved much further by introducing a dynamic model of the pneumatic muscle's fitting equation.

ACKNOWLEDGMENT

This work was supported by KAKENHI: Grant-in-Aid for Scientific Research (B) 21300202.